\documentstyle[12pt]{article}
\def\da{{\dot{\alpha}}}
\def\db{{\dot{\beta}}}
\def\a{\alpha}
\def\b{\beta}
\def\rar{\rightarrow}
\def\dg{{\dagger}}
\def\le{\left(}
\def\ri{\right)}
\def\t{\theta}
\def\bt{{\bar{\theta}}}
\def\ve{\varepsilon}
\def\Db{{\bar{D}}}
\def\yb{{\bar{y}}}

\def\f12{\frac{1}{2}}
\def\fra1g2{\frac{1}{g^2}}
\def\Tr{{\rm Tr}}
\def\Lac{\Lambda}

\def\dis{\displaystyle}
\def\del{\delta}
\def\G{\Gamma}
\def\F{\Phi}
\def\bF{\bar{\Phi}}
\def\bW{\bar{W}}

\def\yb{\bar{y}}

\def\no{\nonumber}
\def\Vt{\tilde{V}}
\def\Kt{\tilde{K}}

\begin{document}
\begin{titlepage}
\flushright{USM-TH-114}\\
\vskip 2cm
\begin{center}
{\Large \bf The solution to Slavnov--Taylor identities in \\
\vspace{3mm} a general four dimensional supersymmetric theory}
\vskip 1cm  Igor Kondrashuk \vskip 5mm {\it Departamento de
Fis\'\i ca, Univeridad T\'ecnica Federico Santa Mar\'\i a, \\
Avenida Espa\~{n}a 1680, Casilla 110-V, Valpara\'\i so, Chile}
\end{center}
\vskip 20mm
\begin{abstract}
A solution to Slavnov--Taylor identities in a general four
dimensional N=1 supersymmetric Yang-Mills theory containing
arbitrary matter superfields is proposed. The solution proposed
appears just a simple generalization of an analogous solution in
the pure supersymmetric Yang-Mills theory.
\end{abstract}
\vskip 1cm
\begin{center}
\end{center}
\end{titlepage}

\section{Introduction}

Recently the solution to Slavnov--Taylor (ST) identities has been proposed
for supersymmetric N=1 theories without matter \cite{jhep}. In this paper under
the effective action we mean what is called the Legendre transformation of
$ln~Z[J]$ with respect to $J,$ where $Z[J]$ is defined in the traditional way
as the path integral in the presence of the external source $J.$ The notation used
for four dimensional  supersymmetry are the same as those have been used in Ref.
\cite{jp}.

In the previous paper \cite{jhep} we have described the procedure of deriving
the solution to the Slavnov--Taylor identities for the case of
the four dimensional Yang--Mills pure gauge theory with softly broken supersymmetry.
In this section we make a short review
of the method proposed there to proceed further with matter superfields.
We consider for simplicity the case of
the unbroken (rigid) supersymmetric N=1 Yang--Mills theory. The only difference
from the softly broken case is that the ``coupling'' is really
a coupling constant but not the external background superfield.

 The path integral describing the quantum theory is defined as
\begin{eqnarray}
& \dis{Z[J,\eta,\bar{\eta},\rho,\bar{\rho},K,L,\bar{L}] = \int~
dV~dc~d\bar{c}~db~d\bar{b}~\exp i}\left[\dis{S}
\right.  \label{pathR}\\
& \left. + \dis{2~\Tr\le\int~d^8z~JV + i\int~d^6y~\eta c
+ i\int~d^6\yb~\bar{\eta}\bar{c} + i\int~d^6y~\rho b
+ i\int~d^6\yb~\bar{\rho}\bar{b}\ri} \right. \no\\
& \left. + \dis{2~\Tr\le i\int~d^8z~K\del_{\bar{c},c}V + \int~d^6y~Lc^2 +
\int~d^6\yb~\bar{L}\bar{c}^2 \ri }\right]. \no
\end{eqnarray}
The third term in the brackets is BRST invariant since the external
superfields $K$ and $L$ are BRST invariant by definition. All fields
in the path integral are in the adjoint representation of the gauge group.
For the sake of brevity we use the following definition for measures in
the superspace
\begin{eqnarray*}
d^8z \equiv d^4x~d^2\t~d^2\bt,~~~ d^6y \equiv d^4y~d^2\t,~~~
d^6\yb \equiv d^4\yb~d^2\bt.
\end{eqnarray*}
The total gauge part of the classical action is
\begin{eqnarray}
& S = \dis{\int~d^6y
~\fra1g2\frac{1}{2^7}\Tr~ W_\a W^\a  + \int~d^6\yb
~\fra1g2\frac{1}{2^7}\Tr~ \bW^\da \bW_\da } \no \\ &  +
\dis{\int~d^8z ~\frac{1}{16}\frac{1}{\a}\Tr~\le\Db^2
V\ri\le D^2V\ri}
  \label{SR} \\
& + \dis{\int~d^6y~\frac{i}{2}\Tr~b~\Db^2
\del_{\bar{c},c}V  + \int~d^6\yb~\frac{i}{2}\Tr~
\bar{b}~D^2\del_{\bar{c},c} V.} \no
\end{eqnarray}
where $b$ and $\bar{b}$ are the antighost chiral and antichiral superfields,
and $c$ and $\bar{c}$ are the ghost chiral and antichiral superfields.
Such a choice of the gauge fixing and the ghost terms means that we fix
the gauge arbitrariness by imposing the condition
\begin{eqnarray*}
D^2 V(x,\t,\bt) = \bar{f}(\yb,\bt),  ~~~  \Db^2 V(x,\t,\bt) = f(y,\t),
\end{eqnarray*}
where  $\bar{f}$ and $f$ are arbitrary antichiral and chiral functions,
respectively.

Having shifted the antighost superfields $b$ and $\bar{b}$ by
arbitrary chiral and antichiral superfields respectively, or,
having made the change of variables in the path integral
(\ref{pathR}) which are the BRST transformations  of the total
gauge action (\ref{SR}), we obtain ghost equations \cite{Piguet}
\begin{eqnarray}
\frac{\del \G}{\del\bar{b}}  - \frac{1}{4}D^2
\frac{\del \G}{\del K} = 0,  ~~~
\frac{\del \G}{\del b} - \frac{1}{4}\Db^2
\frac{\del \G}{\del K} = 0 \label{ghostE}
\end{eqnarray}
and ST identities
\begin{eqnarray}
& \Tr\left[\dis{\int~d^8z~\frac{\del \G}{\del V}\frac{\del \G}{\del K}
 - i \int~d^6y~\frac{\del \G}{\del c}\frac{\del \G}{\del L}
+ i \int~d^6\yb~\frac{\del \G}{\del \bar{c}}\frac{\del \G}{\del \bar{L}} -
\int~d^6y~\frac{\del \G}{\del b}\le\frac{1}{32}\frac{1}{\a}\Db^2D^2V\ri}
\right. \no \\
& \left. - \dis{\int~d^6\yb~\frac{\del \G}{\del
\bar{b}}\le\frac{1}{32}\frac{1}{\a}D^2\Db^2V\ri} \right]=0,  \label{STr}
\end{eqnarray}
respectively. All Grassmannian derivatives in this paper are left
derivatives by definition. From the effective action $\G$ we can
extract a 2-point ghost proper correlator  $G^{(2)}(z-z'),$
\begin{eqnarray*}
\le G^{(2)} \ri ^{\dg} = G^{(2)},
\end{eqnarray*}
and a 2-point connected ghost correlator ${}^{(-1)}G^{(2)}(z_2 -z'),$
which is related to the previous one in the following way
\begin{eqnarray*}
\int d^8z'~G^{(2)}(z_1 - z')~{}^{(-1)}G^{(2)}(z_2 - z') =
\del^{(8)}(z_2 - z_1).
\end{eqnarray*}
In Ref. \cite{jhep} it has been shown that the unique solution to ST identities
(\ref{STr}) has the following form
\begin{eqnarray}
& \dis{\G  = \int~d^6y~f(g^2)~ \Tr~ W_\a\le V \star{}^{(-1)}
G^{(2)}\ri W^\a\le V \star {}^{(-1)}G^{(2)}\ri + {\rm H.c.}} \no\\
& + \dis{\int~d^6y~f_2(g^2)~\Tr~}\le \dis{W_\a\le V
\star{}^{(-1)} G^{(2)}\ri W^\a\le V \star {}^{(-1)}G^{(2)}\ri}
\right. \no\\ &  \left. \dis{W_\b \le V \star
{}^{(-1)}G^{(2)}\ri W^\b \le V \star {}^{(-1)}G^{(2)}\ri} \ri
  + {\rm H.c.} + \dots \no \\
& + \dis{\int~d^8z~\frac{1}{32}\frac{1}{\a}\Tr V\le D^2\Db^2 +
\Db^2D^2\ri V} \label{Gfull}\\
& + \dis{\int~
d^8z~d^8z'~2i~\Tr \le b(z) +\bar{b}(z)\ri
~G^{(2)}(z-z')~\del_{\bar{c},c} \le V \star{}^{(-1)} G^{(2)}\ri
(z')  } \no\\ & + \dis{\int~d^8z~2i~\Tr~\le K\star G^{(2)}\ri (z)
~\del_{\bar{c},c}\le V \star{}^{(-1)} G^{(2)}\ri (z)} \no\\
& + \dis{\int~d^6y~2\Tr~Lc^2 + \int~d^6\yb~2\Tr~\bar{L}\bar{c}^2}. \no
\end{eqnarray}
Here we have introduced for the brevity the following notation  for the
integral convolutions
\begin{eqnarray}
\dis{ V \star{}^{(-1)} G^{(2)}  = \int~d^8z'~V(z')~{}^{(-1)}G^{(2)}(z-z')}.
\label{conv}
\end{eqnarray}
One can check that this action satisfies to the identity
(\ref{STr}). Some comments are necessary here. First, $f(g^2)$
and $f_2(g^2)$ are functions of the gauge coupling. Second,
the multidots $\dots$ denote terms of higher orders in
$W_\a$ invariant with respect to chiral rotations
\begin{eqnarray*}
\dis{W_\a \rar  e^{-\Lac} W_\a e^{\Lac}}.
\end{eqnarray*}

\section{Including the matter}

In this section we include the matter in the ST identity (\ref{STr}).
The Lagrangian with the matter takes the form
\begin{eqnarray*}
& S = \dis{\int~d^6y~\fra1g2\frac{1}{2^7}\Tr~ W_\a W^\a  +
\int~d^6\yb~\fra1g2\frac{1}{2^7}\Tr~ \bW^\da \bW_\da } \nonumber\\
& + \dis{\int~d^8z~\bF~e^V~\F}  \no\\
& + \dis{\int~d^6y~\left[Y^{ijk}\F_i\F_j\F_k + M^{ij}\F_i\F_j \right]}
+ \dis{\int~d^6\yb~\left[\bar{Y}_{ijk}\bF^i\bF^j\bF^k \no
+ \bar{M}_{ij}\bF^i\bF^j \right]}.
\end{eqnarray*}
We take the same gauge fixing condition as we did in the previous
section and the total gauge part is coinciding with (\ref{SR}).
The matter superfield $\F$ is in an appropriate, in general, reducible
representation of the gauge group with a set of irreducible
representations.

The path integral describing the quantum theory is defined as
\begin{eqnarray}
& \dis{Z[J,\eta,\bar{\eta},\rho,\bar{\rho},j,\bar{j},K,L,\bar{L},k,\bar{k}] = \int~
dV~dc~d\bar{c}~db~d\bar{b}~d\F~d\bF~\exp i}\left[\dis{S}
\right.  \no \\
& \left. + \dis{2~\Tr\le\int~d^8z~JV + i\int~d^6y~\eta c
+ i\int~d^6\yb~\bar{\eta}\bar{c} + i\int~d^6y~\rho b
+ i\int~d^6\yb~\bar{\rho}\bar{b}\ri} \right. \no\\
& \left. + \dis{\le\int~d^6y~\F~j + \int~d^6\yb~\bar{j}~\bF\ri }
    \right. \label{pathRM}\\
& \left. + \dis{2~\Tr\le i\int~d^8z~K\del_{\bar{c},c}V + \int~d^6y~Lc^2 +
\int~d^6\yb~\bar{L}\bar{c}^2 \ri }\right. \no\\
& \left. + \dis{\int~d^6y~k~c~\F + \int~d^6\yb~\bF~\bar{c}~\bar{k}} \right], \no
\end{eqnarray}
where
\begin{eqnarray}
& S = \dis{\int~d^6y
~\fra1g2\frac{1}{2^7}\Tr~ W_\a W^\a  + \int~d^6\yb
~\fra1g2\frac{1}{2^7}\Tr~ \bW^\da \bW_\da } \no \\ &  +
\dis{\int~d^8z ~\frac{1}{16}\frac{1}{\a}\Tr~\le\Db^2
V\ri\le D^2V\ri} \no \\
& + \dis{\int~d^6y~\frac{i}{2}\Tr~b~\Db^2
\del_{\bar{c},c}V  + \int~d^6\yb~\frac{i}{2}\Tr~
\bar{b}~D^2\del_{\bar{c},c} V} \label{SRM} \\
& + \dis{\int~d^8z~\bF~e^V~\F}  \no\\
& + \dis{\int~d^6y~\left[Y^{ijk}\F_i\F_j\F_k + M^{ij}\F_i\F_j \right]}
+ \dis{\int~d^6\yb~\left[\bar{Y}_{ijk}\bF^i\bF^j\bF^k \no
+ \bar{M}_{ij}\bF^i\bF^j \right]}.
\end{eqnarray}

The action (\ref{SRM}) is invariant under the BRST symmetry \cite{Piguet},
\begin{eqnarray}
& \dis{e^V \rar e^{i\large{\bar{c}\ve}} e^V e^{ic\ve}},  \no \\
& \dis{c \rar c + ic^2\ve}, ~~~ \dis{\bar{c} \rar \bar{c} - i\bar{c}^2\ve}, \no \\
& \dis{\del b = \frac{1}{32}\frac{1}{\a}\le\Db^2D^2V\ri\ve}, ~~~
  \dis{\del \bar{b} = \frac{1}{32}\frac{1}{\a}\le D^2\Db^2V\ri\ve}, \no\\
& \dis{\del \F = -ic\ve~\F,~~~\del \bF = -\bF~i\bar{c}\ve}  \label{BRST}
\end{eqnarray}
with an Hermitian Grassmannian parameter $\ve$, $\ve^\dg = \ve.$
The external sources $k$, $\bar{k}$ of the BRST transformations of chiral
multiplets are BRST invariant by definition, so the last two lines in the
eq. (\ref{pathRM}) are BRST invariant.

The effective action $\G$ is related to $W = i~ln~Z$ by the Legendre
transformation
\begin{eqnarray}
& \dis{V \equiv - \frac{\del W}{\del J}, ~~ ic  \equiv - \frac{\del W}{\del \eta}, ~~
  i\bar{c}  \equiv - \frac{\del W}{\del \bar{\eta}}, ~~
  ib \equiv - \frac{\del W}{\del \rho}, ~~i\bar{b} \equiv -
  \frac{\del W}{\del \bar{\rho}}}, \no\\
& \dis{\F \equiv - \frac{\del W}{\del j}, ~~ \bF \equiv -
\frac{\del W}{\del \bar{j}},}
      \label{defphi} \\
& \dis{\G = - W - 2~\Tr\le \int~d^8z~JV + \int~d^6y~i\eta c
+ \int~d^6\yb~i\bar{\eta}\bar{c} + \int~d^6y~i\rho b +
  \int~d^6\yb~i\bar{\rho}\bar{b}\ri} \no\\
& - \dis{\int~d^6y~\F~j - \int~d^6\yb~\bar{j}~\bF}
  \equiv \dis{- W - 2~\Tr\le X\phi\ri - \le\F~j\ri - \le\bar{j}~\bF\ri},
                                      \label{Legendre} \\
&  \dis{\le X\phi\ri \equiv i^{G(k)}\le X^{k}\phi^{k}\ri,}  \no\\
& \dis{X \equiv \le J,\eta,\bar{\eta},\rho,\bar{\rho}\ri,  ~~~   \phi \equiv
  \le V,c,\bar{c},b,\bar{b} \ri}. \no
\end{eqnarray}
where $G(k) = 0$ if $\phi^{k}$ is the Bose superfield and $G(k) = 1$
if $\phi^{k}$ is the Fermi superfield. Iteratively all equations (\ref{defphi})
can be reversed,
\begin{eqnarray*}
& X = X[\phi,\F,\bF,K,L,\bar{L},k,\bar{k}], \\
& j = j[\phi,\F,\bF,K,L,\bar{L},k,\bar{k}], \\
& \bar{j} = \bar{j}[\phi,\F,\bF,K,L,\bar{L},k,\bar{k}],
\end{eqnarray*}
and the effective action is defined in terms of new variables,
$\G = \G[\phi,\F,\bF,K,L,\bar{L},k,\bar{k}].$
Hence, the following equalities take place
\begin{eqnarray}
& \dis{\frac{\del \G}{\del V} = - J, ~~~\frac{\del \G}{\del \F} = -j, ~~~
\frac{\del \G}{\del \bF} = -\bar{j}}, \no\\
& \dis{\frac{\del \G}{\del K} = - \frac{\del W}{\del K}, ~~~
\frac{\del \G}{\del k} = - \frac{\del W}{\del k}, ~~~
\frac{\del \G}{\del \bar{k}} = - \frac{\del W}{\del \bar{k}}}, \label{GW}\\
& \dis{\frac{\del \G}{\del c} = i\eta, ~~ \frac{\del \G}{\del\bar{c}} = i\bar{\eta}, ~~
\frac{\del \G}{\del b} = i\rho, ~~ \frac{\del \G}{\del\bar{b}} = i\bar{\rho},  ~~
\frac{\del \G}{\del L} = - \frac{\del W}{\del L}, ~~
\frac{\del \G}{\del \bar{L}} = - \frac{\del W}{\del \bar{L}}}. \no
\end{eqnarray}

If the change of fields (\ref{BRST}) in the path integral (\ref{pathRM}) is made
one obtains the Slavnov--Taylor identity as the result of invariance of the
integral (\ref{pathRM}) under a change of variables,
\begin{eqnarray*}
& \Tr\left[\dis{\int~d^8z~J\frac{\del}{\del K} -
\int~d^6y~i\eta\le\frac{1}{i}\frac{\del}{\del L}\ri +
\int~d^6\yb~i\bar{\eta}\le\frac{1}{i}\frac{\del}{\del \bar{L}}\ri +
\int~d^6y~i\rho\le\frac{1}{32}\frac{1}{\a}\Db^2D^2\frac{\del}{\del J}\ri} \right. \\
& \left. +
\dis{\int~d^6\yb~i\bar{\rho}
\le\frac{1}{32}\frac{1}{\a}D^2\Db^2\frac{\del}{\del J}\ri}\right]W
+ \dis{\left[\int~d^6y~ j~\le\frac{1}{i}\frac{\del }{\del k}\ri -
\int~d^6\yb~\le\frac{1}{i}\frac{\del }{\del \bar{k}}\ri ~\bar{j}\right]W} =0,
\end{eqnarray*}
or, taking into account the relations (\ref{GW}), we have
\cite{Piguet}
\begin{eqnarray}
& \Tr\left[\dis{\int~d^8z~\frac{\del \G}{\del V}\frac{\del \G}{\del K}
 - i \int~d^6y~\frac{\del \G}{\del c}\frac{\del \G}{\del L}
+ i \int~d^6\yb~\frac{\del \G}{\del \bar{c}}\frac{\del \G}{\del
\bar{L}} - \int~d^6y~\frac{\del \G}{\del
b}\le\frac{1}{32}\frac{1}{\a}\Db^2D^2V\ri} \right. \no \\ & \left.
- \dis{\int~d^6\yb~\frac{\del \G}{\del
\bar{b}}\le\frac{1}{32}\frac{1}{\a}D^2\Db^2V\ri} \right] -
\dis{i~\int~~d^6y~\frac{\del \G}{\del \F}~\frac{\del \G}{\del k} +
i~\int~d^6\yb~\frac{\del \G}{\del \bar{k}}~\frac{\del \G}{\del
\bF} = 0}. \label{STrM}
\end{eqnarray}

\section{Solution to ST identities}

In this section we present a general solution to the identity
(\ref{STrM}). The problem is to find a functional $\G$ of the variables
$\phi,\F,\bF,K,L,\bar{L},k,\bar{k}$ that satisfies to the ST identity
(\ref{STrM}). There are several steps in the procedure of searching the
solution. First, the gauge part of the solution that corresponds
to the part of  $\G$ with zero values of $\F,\bF,k,\bar{k}$
is already found. It is the expression (\ref{Gfull}). Second,
we suppose according to the no-renormalization theorem for
supersymmetric theories that the chiral potential
does not obtain any correction in the effective action in
comparison with the classical action \cite{West}. Thus, the form of the
superpotential is
\begin{eqnarray}
& \dis{\int~d^6y~\left[Y^{ijk}\F_i\F_j\F_k + M^{ij}\F_i\F_j \right]}
+ \dis{\int~d^6\yb~\left[\bar{Y}_{ijk}\bF^i\bF^j\bF^k \no
+ \bar{M}_{ij}\bF^i\bF^j \right]} \label{poten}\\
& + \dis{\int~d^6y~2\Tr~Lc^2 + \int~d^6\yb~2~\Tr~\bar{L}\bar{c}^2
+ \int~d^6y~k~c~\F + \int~d^6\yb~\bF~\bar{c}~\bar{k}}.
\end{eqnarray}
As to the gauge-matter interaction part, it is easy to
guess the most general form of a term of this part of $\G$
which is
\begin{eqnarray}
\bF~\bar{\nabla}^\da \dots \bar{\nabla}^\db~e^{\Vt}~\nabla_\a \dots \nabla_\b~
e^{-\Vt}~\bar{\nabla}_\da \dots \bar{\nabla}_\db~
e^{\Vt}~\nabla^\a \dots \nabla^\b~\F, \label{M}
\end{eqnarray}
where $\Vt$ means
\begin{eqnarray*}
\dis{V(z) = \int~d^8z'~\Vt(z')~G^{(2)}(z-z')}.\no
\end{eqnarray*}

The argument why it must be so is very short. Indeed, as it has been
shown in Ref. \cite{jhep} after the change of variables
\begin{eqnarray*}
& \dis{K(z) = \int~d^8z'~\tilde{K}(z')~{}^{(-1)}G^{(2)}(z-z')}, \\
& \dis{V(z) = \int~d^8z'~\tilde{V}(z')~G^{(2)}(z-z')}
\end{eqnarray*}
in the effective action functional $\G$ the identity (\ref{STr})
recovers the functional $\G$ in a gauge invariant manner with
respect to new variables $\Vt$ and $\Kt$.  It means that in terms
of these new variables the functional $\G$ (\ref{Gfull}) has a
gauge invariant structure for the field $\Vt$ and all the action
(\ref{Gfull}) as whole is BRST invariant in terms of the new
variables. Having included the matter, one can see that the ST
identity (\ref{STrM}) recovers the effective action in a gauge
invariant manner in terms of $\Vt$ and $\Kt$ too, with a gauge
transformation of matter fields defined by the terms in the last
line in eq. (\ref{poten}). Apparently, the structure (\ref{M}) is
gauge invariant if we take into account the definitions
\cite{GGRS} of the covariant derivatives
\begin{eqnarray*}
\nabla_\a = e^{-\Vt}~D_\a~e^{\Vt}, ~~~ \bar{\nabla}_\db = e^{\Vt}~\Db_\db~e^{-\Vt}.
\end{eqnarray*}
Any convolution of the spinor indices in (\ref{M})  is allowed.
For example, one can obtain
\begin{eqnarray*}
\bF~\bar{\nabla}^2~e^{\Vt}~\nabla^2~e^{-\Vt}~\bar{\nabla}^2~e^{\Vt}~\nabla^2~\F,
\end{eqnarray*}
or
\begin{eqnarray*}
\bF~\bar{\nabla}^\da~e^{\Vt}~\nabla_\a~e^{-\Vt}~\bar{\nabla}_\da~e^{\Vt}~
\nabla^\a~\F.
\end{eqnarray*}
Thus, all the structures like (\ref{M}), their degrees and
convolutions in Lorentz indices can be present in the effective
action. The coefficients before such structures can not be fixed
in the framework of the approach proposed here. At the same time,
this approach allows to restrict the functional structure of the
effective action.

\section{Conclusions}

To conclude we would like to combine the pure gauge part (\ref{Gfull}),
the superpotential (\ref{poten}) and the gauge-matter interaction
part (\ref{M}) together in order to obtain the total answer for
the solution to the identities (\ref{STrM}). Thus, the final result is
\begin{eqnarray}
& \dis{\G  = \int~d^6y~f(g^2,Y,\bar{Y})~ \Tr~ W_\a\le V \star{}^{(-1)}
G^{(2)}\ri W^\a\le V \star {}^{(-1)}G^{(2)}\ri + {\rm H.c.}} \no\\
& + \dis{\int~d^6y~f_2(g^2,Y,\bar{Y})~\Tr~}\le \dis{W_\a\le V
\star{}^{(-1)} G^{(2)}\ri W^\a\le V \star {}^{(-1)}G^{(2)}\ri}
\right. \no\\ &  \left. \dis{W_\b \le V \star
{}^{(-1)}G^{(2)}\ri W^\b \le V \star {}^{(-1)}G^{(2)}\ri} \ri
  + {\rm H.c.} + \dots \no \\
& + \dis{\int~d^8z~\frac{1}{32}\frac{1}{\a}\Tr V\le D^2\Db^2 +
\Db^2D^2\ri V} \no\\
& + \dis{\int~
d^8z~d^8z'~2i~\Tr \le b(z) +\bar{b}(z)\ri
~G^{(2)}(z-z')~\del_{\bar{c},c} \le V \star{}^{(-1)} G^{(2)}\ri
(z')  } \no\\ & + \dis{\int~d^8z~2i~\Tr~\le K\star G^{(2)}\ri (z)
~\del_{\bar{c},c}\le V \star{}^{(-1)} G^{(2)}\ri (z)} \no\\
& + \dis{\int~d^6y~\left[Y^{ijk}\F_i\F_j\F_k + M^{ij}\F_i\F_j \right]}
+ \dis{\int~d^6\yb~\left[\bar{Y}_{ijk}\bF^i\bF^j\bF^k \no
+ \bar{M}_{ij}\bF^i\bF^j \right]} \no\\
& + \dis{\int~d^6y~2\Tr~Lc^2 + \int~d^6\yb~2\Tr~\bar{L}\bar{c}^2
+ \int~d^6y~k~c~\F + \int~d^6\yb~\bF~\bar{c}~\bar{k}} \no\\
& + \dis{\int~d^8z~\sum~F(g^2,Y,\bar{Y})~\bF(z)~\bar{\nabla}^\da \dots \bar{\nabla}^\db~
e^{ V \star {}^{(-1)}G^{(2)}}~
  \nabla_\a \dots \nabla_\b~
  e^{- V \star {}^{(-1)}G^{(2)}}~\bar{\nabla}_\da \dots} \no \\
& \dis{\dots \bar{\nabla}_\db~
  e^{ V \star {}^{(-1)}G^{(2)}}~\nabla^\a \dots \nabla^\b~\F(z).}  \no
\end{eqnarray}
Some comments about the last line of this expression which is
responsible for gauge-matter interaction are necessary. First,
we have taken the symbolic sum over all possible terms (\ref{M}) allowed by
Slavnov--Taylor identities. Second, $F(g^2,Y,\bar{Y})$ are some numbers
which depend on couplings of the theory and can not be fixed
in the framework of our approach. Third, in the covariant derivatives
we have made the transformation from the variable $\Vt$ to the initial
variable $V,$ that is
\begin{eqnarray*}
& \dis{\Vt(z) = \int~d^8z'~V(z')~ {}^{(-1)}G^{(2)}(z-z')}, \\
& \dis{\nabla_\a = e^{-V \star{}^{(-1)}G^{(2)}}~D_\a~e^{V \star{}^{(-1)}
   G^{(2)}}}.
\end{eqnarray*}
One can check by a direct substitution that the result obtained
here is the solution to the ST identity (\ref{STrM}). In case of
Abelian vector superfield the ghost 2-point function turns out to
be trivial $\del$-function in the superspace, $\del(z-z'),$ so
that the correspondence with Abelian case is obvious. Traditional
background technique \cite{GGRS} usually used for calculations of
perturbative characteristics in a general supersymmetric
Yang--Mills theory  can be also reproduced in the framework of the
approach developed here with small modifications of the procedure.
Thus, the proposed method could be useful in studying the  quantum
behaviour of a general supersymmetric quantum field theory with
arbitrary matter contents.

\vskip 3mm
\noindent {\large{\bf{Acknowledgements}}}
\vskip 3mm

 I am grateful to Ivan Schmidt for useful discussions. This work is
supported by the Programa MECESUP FSM9901 of the Ministry of
Education (Chile) and also by Conicyt (Chile) under grant 8000017.


\begin{thebibliography}{99}
\bibitem{jhep} I. Kondrashuk, JHEP 0011:034(2000)
\bibitem{jp} I. Kondrashuk,  J.Phys.A33(2000)6399
\bibitem{Piguet} O. Piguet, "Supersymmetry, supercurrent, and scale invariance",
                hep-th/9611003
\bibitem{GGRS} J. Gates Jr., M.T. Grisaru, M. Ro\v{c}ek, W. Siegel ``One Thousand
                and One Lessons in Supersymmetry'' Benjamin/Cummings, 1983.
\bibitem{West} P. West, "Introduction to Supersymmetry and Supergravity",
                World Scientific, 1986.
\end{thebibliography}
\end{document}